\documentclass[aps,prl,reprint,superscriptaddress,,nofootinbib,showpacs]{revtex4-1}
\usepackage{amsmath,amssymb,bm,float,mathrsfs}
\usepackage{graphicx}
\usepackage{dcolumn}
\usepackage{color}
\usepackage{hyperref}
\usepackage{comment}
\usepackage{fancybox}

  \newcommand{\beq}{\begin{equation}}
  \newcommand{\eeq}{\end{equation}}
  \newcommand{\al}[1]{\begin{align} #1 \end{align}}
  \newcommand{\bi}{\begin{itemize}}
  \newcommand{\ei}{\end{itemize}}
  \def\dd{\mathrm{d}}

  \def\pd{\partial}
  \newcommand{\ave}[1]{\left\langle #1 \right\rangle}

\begin{document}

\title{Observational signatures of dark energy produced
in an ancestor vacuum:\\
Forecast for galaxy surveys}

\author{Daisuke Yamauchi}
\email[Email: ]{yamauchi@jindai.jp}
\affiliation{
Faculty of Engineering, Kanagawa University, Kanagawa, 221-8686, Japan
}

\author{Hajime Aoki}
\email[Email: ]{haoki@cc.saga-u.ac.jp}
\affiliation{
Department of Physics, Saga University, Saga 840-8502, Japan
}

\author{Satoshi Iso}
\email[Email: ]{iso@post.kek.jp}
\affiliation{
Theory Center,
 High Energy Accelerator Research Organization (KEK),
 and Graduate University for Advanced Studies (SOKENDAI),
Ibaraki 305-0801, Japan
}

\author{Da-Shin Lee}
\email[Email: ]{dslee@gms.ndhu.edu.tw}
\affiliation{
Department of Physics, National Dong-Hwa University,
 Hualien 97401, Taiwan, R.O.C.
}

\author{Yasuhiro Sekino}
\email[Email: ]{ysekino@la.takushoku-u.ac.jp}
\affiliation{
Department of Liberal Arts and Sciences,
Faculty of Engineering,
Takushoku University, Tokyo 193-0985, Japan
}

\author{Chen-Pin Yeh}
\email[Email: ]{chenpinyeh@gmail.com}
\affiliation{
Department of Physics, National Dong-Hwa University,
 Hualien 97401, Taiwan, R.O.C.
}

\begin{abstract}
We study observational consequences of the model for dark energy
proposed in \cite{Aoki:2017scu}. We 
assume our universe has been created by bubble nucleation, 
and consider quantum fluctuations of an ultralight scalar field. 
Residual effects of fluctuations generated in an ancestor vacuum 
(de Sitter space in which the bubble was formed) 
is interpreted as dark energy. Its equation of state parameter 
$w_{\rm DE}(z)$ has a characteristic form, approaching
$-1$ in the future, but $-1/3$ in the past. A novel feature
of our model is that dark energy effectively increases
the magnitude of the negative spatial curvature
in the evolution of the Hubble parameter,
though it does not alter the definition of 
the angular diameter distance.
We perform Fisher analysis and forecast the constraints 
for our model from future galaxy surveys by Square Kilometre 
Array and Euclid. Due to degeneracy between dark energy 
and the spatial curvature, galaxy surveys alone can
determine these parameters only for optimistic choices
of their values, but combination with other independent 
observations, such as CMB, will greatly improve the chance 
of determining them.

\end{abstract}






\maketitle

\paragraph*{Introduction.---}
It is widely accepted that the expansion of the
present universe is accelerating. 
The first clear evidence for acceleration has been provided
by observations of supernovae (SNe) of type 
Ia~\cite{Perlmutter:1998np,Riess:1998cb}.
Strong support has been given by the fact that other
independent phenomena such as cosmic microwave 
background (CMB), baryon acoustic oscillation (BAO), 
etc., consistently suggest cosmic acceleration
(see e.g., \cite{Weinberg:2012es} for a review).
The source of the accelerating expansion is attributed 
to dark energy of unknown origin, which has
the equation of state (EoS) parameter $w_{\rm DE}=p/\rho$ 
close to $-1$, contributing about 68\% of the critical 
density~\cite{Ade:2015xua, Ade:2015rim}. 
The next generation observations are expected
to determine $w_{\rm DE}$ to a percent level, and also 
its derivative
with respect to the scale factor to a
10 percent level~\cite{Weinberg:2012es}.

At a time of great observational developments, an important
direction of research would be to construct a theoretically 
motivated model of dark energy, and have it tested by 
observations. There are models which describe
dark energy (see \cite{Amendola:2015ksp} for a review), 
such as quintessence (scalar field slowly rolling down a 
potential) and modified gravity, but they typically do not
explain its origin. It remains a
mystery why its energy density is extremely small 
compared to the fundamental scale, 
$\rho_{\rm DE}\sim 10^{-122}M_{\rm P}^4$.

\begin{figure}[!htb]
\center
\includegraphics [scale=.18]
 {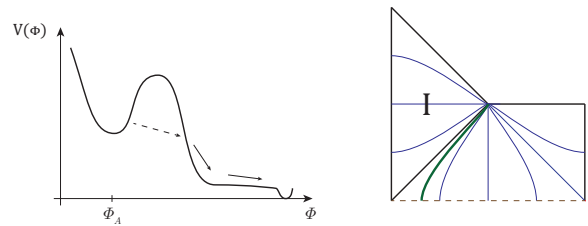} \caption{Left panel: A potential with 
a metastable de Sitter vacuum. 
Right panel: Penrose diagram for a bubble nucleated in 
de Sitter space. The horizontal line at the bottom 
represents a global slice in de Sitter
space at the time of bubble nucleation. 
The green line is the bubble wall;
on its left (right) is the true (ancestor) vacuum. 
Our universe is an open FLRW universe inside the
bubble, represented as Region~I.
}
\label{Fig-Penrose}
\end{figure}

In the previous paper~\cite{Aoki:2017scu}, five of the present 
authors proposed a model for dark energy, partially motivated
by string landscape. Assuming our universe has been
created by bubble nucleation from a metastable
de Sitter space (the ``ancestor vacuum''), 
residual effects of quantum 
fluctuations generated in 
the ancestor vacuum has been interpreted as the source 
of dark energy. Its equation of state parameter $w_{\rm DE}(z)$
has a characteristic form as a function of the redshift $z$.  
The purpose of the present Letter is to assess the possibility 
of observational tests of this model. 
Observations that we consider to be particularly promising 
are galaxy surveys by Square 
Kilometre Array (SKA)~\cite{SKA} 
and Euclid~\cite{Euclid}, planned to start operating
in the early 2020's.
We will perform Fisher analysis and forecast the constraints
for our model\footnote{%
Our work focuses on the test of a particular model, and
is complementary to the attempts at 
``model independent'' forecasts~\cite{Kohri:2016bqx}.
The authors of \cite{Kohri:2016bqx}
introduce as many parameters 
as to be considered general,
and study emissions from intergalactic medium at 
$6.3\lesssim z\lesssim 20$, while we consider a
minimal set of parameters and study
emissions from galaxies themselves at $z\lesssim 2$.}.

\paragraph*{The theoretical setup.---}
We consider bubble nucleation
due to quantum tunneling, which occurs
e.g., in a potential for a scalar field $\Phi$
shown in FIG.~\ref{Fig-Penrose}, left panel. 
The Hubble parameter of the ancestor vacuum will be
denoted $H_A$. Our universe is inside the bubble, 
represented as Region~I in 
FIG.~\ref{Fig-Penrose}, right panel. It should have negative 
spatial curvature~\cite{Coleman:1980aw}. After bubble nucleation,
the ordinary inflation with the Hubble parameter 
$H_I(\ll H_A)$ is assumed to occur.

We consider a scalar field $\phi$ which is different
from the tunneling field\footnote{%
The tunneling field does not have a supercurvature
mode~\cite{Garriga:1998he}
which will give the residual effect at late
times (as explained below), 
and cannot serve as a source for dark energy.
Gravitons~\cite{Tanaka:1997kq} and 
vector fields~\cite{Yamauchi:2014saa} also do not have one.} 
$\Phi$ and has zero expectation value. 
We assume the field $\phi$ to have mass $m_A$ before tunneling, 
and $m_0$ after tunneling, where $m_A$ and $m_0$ could be 
different.
The latter is assumed to satisfy
$m_0\lesssim H_0$ where $H_0\sim 10^{-33} {\rm eV}$ 
is the present Hubble parameter. 
A candidate for such an ultralight field is one of the
axion-like fields, expected to exist in string 
compactifications\footnote{%
Extremely small mass
$m_0\sim H_0$ is considered to be possible because
mass of a string axion arises 
typically due to instanton effects, and is
exponentially sensitive to
the instanton action~\cite{Arvanitaki:2009fg}.}~\cite{Arvanitaki:2009fg}.

In \cite{Aoki:2017scu}, the contribution from the
field $\phi$ to the vacuum expectation value 
 of the energy-momentum tensor
$\langle T_{\mu\nu} \rangle$ has been computed, carefully 
taking into account the effect of
the ancestor vacuum. In the free-field approximation,
the energy-momentum tensor is quadratic in $\phi$,
thus $\langle T_{\mu\nu} \rangle$ can
be obtained by taking the coincident-point limit of the
two-point function $\langle \phi\phi\rangle$ computed by 
the method developed in\footnote{%
For studies of the CMB in this framework, see 
e.g.~\cite{Garriga:1998he, Linde:1999wv, Yamauchi:2011qq,
Sugimura:2012kr, Sugimura:2011tk}.} \cite{Yamamoto:1996qq, 
Sasaki:1994yt, Freivogel:2006xu}.
We refer the reader to \cite{Aoki:2017scu} for details, 
and give an order-of-magnitude argument here.
In the ancestor vacuum, quantum
fluctuations give rise to the expectation value of the 
field-squared $\langle \phi^2 \rangle\sim H_A^4/m_A^2$,
as in pure de Sitter space (see e.g.,
\cite{Linde:2005ht}).
The field $\phi$ is almost frozen until now, due to the 
assumption $m_0\lesssim H_0$
(and one more condition $\epsilon\ll 1$, 
to be mentioned below). If so,
the energy-momentum tensor is dominated by the mass term,
and takes the form of cosmological constant ($w_{\rm DE}=-1$),
with the magnitude
\begin{equation}
\langle T_{\mu\nu}\rangle \sim m_0^2\langle \phi^2\rangle
g_{\mu\nu}\sim H_{A}^{4}\left({m_{0}\over m_{A}}\right)^{2}g_{\mu\nu}
\,.
\label{Tmn}
\end{equation}
It is not difficult for this to have 
the same order of magnitude as dark energy, 
$\rho_{\rm DE}\sim M_{P}^2H_{0}^{2}$, once we admit
$m_0\sim H_0$: 
We just need 
$m_{A}/H_{A}\sim H_{A}/M_{P}$, i.e., $H_{A}$ being
the geometric mean of $m_{A}$ and $M_{P}$. 

\paragraph*{Difference from quintessence.---}
At the level of the above heuristic argument, it makes no
difference whether $\phi$ is a classical or quantum field. 
However, fully quantum mechanical analysis in 
\cite{Aoki:2017scu} has the following two important 
differences from the classical case. 

First, there is no ambiguity in the initial condition 
for the field $\phi$, unlike in the classical case, in 
which one has to assign e.g., the axion misalignment angle
by hand. Our prediction \eqref{Tmn} 
is unambiguous when $H_A$, $m_A$, $m_0$ are given.
This is a virtue of bubble nucleation, which allows us 
to go past the beginning of the FLRW time and uniquely
determine the vacuum state of a quantum field.

Second, the mode of $\phi$ which gives the dominant contribution
at late times is not strictly homogeneous. It
is an eigenfunction of
the Laplacian on the spatial slice $H^3$ with 
a non-zero eigenvalue. There is a peculiar feature
of modes on a hyperboloid. Normalizable modes decay
exponentially (since the volume grows exponentially)
at large distances; they have eigenvalues 
$\nabla^{2}=-(k^2+1)$ with real $k$. However,
in an open universe created by bubble nucleation, 
there is the so-called supercurvature
mode~\cite{Yamamoto:1996qq, 
Sasaki:1994yt},
which is non-normalizable on $H^{3}$, and has
an eigenvalue with imaginary $k=i(1-\epsilon)$,
i.e., $\nabla^2\sim -2\epsilon$ when $\epsilon \ll 1$. 
The parameter $\epsilon$ is determined by the properties
of the ancestor vacuum, 
\begin{equation}
\epsilon= c_{\epsilon} \left({m_{A}\over H_{A}}\right)^2,
\end{equation}
when $m_A/H_A\ll 1$, 
with an order-one coefficient $c_{\epsilon}$ which 
depends on the size of the
critical bubble\footnote{%
The dependence on the bubble size is not very strong: 
$c_{\epsilon}=1/3$ in the small bubble limit, 
and $c_{\epsilon}=2/9$ when the bubble occupies half
of the ancestor de Sitter space.}\cite{Aoki:2017scu}.
The supercurvature mode gives rise to long-range correlations
in the open universe, which can be interpreted as the
superhorizon fluctuations in the ancestor vacuum, seen
from the inside of the bubble.

\paragraph*{Observational signature.---}
The energy-momentum tensor for $\phi$ at late times is dominated
by the contribution from the supercurvature mode. 
The spatial derivative term in 
$\langle T_{\mu\nu} \rangle$ gives a 
non-zero contribution,
$\langle (\nabla\phi)^2\rangle=-\langle \phi\nabla^{2}\phi
\rangle\sim 2\epsilon\langle \phi^2\rangle$. 
If $\epsilon\ll 1$ and $m_{0}\ll H_{0}$ (though these
inequalities do not have to be very strong, in practice),  
the time-derivative term is negligible%
~\cite{Aoki:2017scu}. 
Then, the EoS parameter $w_{\rm DE}$ can be obtained 
by simply taking the ratio of $p$ to $\rho$, and becomes
\begin{equation}
w_{\rm DE}= - {{2\over 3}(\epsilon/ R_{c}^2)+m_0^2 a^2\over
  2(\epsilon/R_{c}^2)+m_0^2 a^2}
  =- {1+{2\over 3}\tilde{\epsilon}(1+z)^2
  \over 1+2\tilde{\epsilon}(1+z)^2} \,.
\label{wz}
\end{equation}
where $R_{c}$ is the comoving radius of curvature, which is
constant\footnote{%
The convention in \cite{Aoki:2017scu} was to take $R_c=1$.
Here we take the scale factor at present to be 1, $a_0=1$,
so that $a=1/(1+z)$.}. 
The final expression shows that the functional
form of $w_{\rm DE}(z)$ depends on a single parameter\footnote{%
This corresponds to the $p=2$, $w_0=-1$, $w_1=-1/3$
case in  
\cite{Hannestad:2004cb} (referred to 
``parametrization~II'' in \cite{Kohri:2016bqx}).},
\begin{equation}
  \tilde{\epsilon}\equiv{\epsilon\over (m_0/H_0)^2}\Omega_{{\rm K},0}\equiv\xi\Omega_{\rm K,0}\, ,
\end{equation}
where $\Omega_{{\rm K},0}=1/H_0^2R_c^2$ is the fractional 
energy density of the spatial curvature at present.
At late times, the mass term is dominant in \eqref{wz},
thus $w_{\rm DE}(z)\to -1$. 
At early times\footnote{%
Though we call it early times, we are assuming it to be
later than the time when the supercurvature mode becomes
dominant over the continuous modes.}, 
the spatial derivative term becomes
dominant, thus $w_{\rm DE}(z)\to -1/3$. 
The past asymptotic value $w_{\rm DE}(z)\to -1/3$ is
unlikely to be realized in the ordinary 
inflation: $\langle T_{\mu\nu}\rangle$ in de Sitter 
space with a de Sitter invariant vacuum should have 
$w=-1$; evolution of wave functions after inflation
will give rise to non-zero time derivatives, not only
spatial derivatives. 

We regard the functional form of $w_{\rm DE}(z)$ in 
(\ref{wz}) to be an indication of fluctuations generated
before ordinary inflation\footnote{%
Another possibility for realizing $w_{\rm DE}(z)$ of
\eqref{wz} is to have two independent sources: one 
with $w=-1/3$ (such as cosmic strings) and the other
with $w=-1$
(cosmological constant).}, but in fact, 
this may not be specific to open universe or bubble 
nucleation. If an infrared part of the fluctuations
is enhanced relative to the usual magnitude $H_I$
and is frozen until now, it will 
contribute to the spatial-derivative and mass terms 
of the energy-momentum tensor, giving the EoS parameter 
similar to (\ref{wz}).
This can happen e.g., in a 
double inflation model in \cite{Aoki:2014ita,Aoki:2014dqa}.

Eq.~\eqref{wz} leads to a simple relation between $w_0$
and its derivative $w_a=-dw/da|_{a=1}$:
When $\tilde{\epsilon}\ll 1$, we have 
$w_a=2(w_0+1)=8\tilde{\epsilon}/3$.
This relation could be testable by ongoing 
observations such as Dark Energy Survey~\cite{DES}. 
This will be a first step toward 
testing our model.

Eq.~\eqref{wz} yields the energy density of dark energy
as a function of the redshift,
\al{
	\rho_{\rm DE} (z)
=3M_P^2H_0^2\,\Omega_{\rm \Lambda,0}
\left(1+2\widetilde\epsilon (1+z)^2\right)
	\,.
\label{rhoDE}
}
The mass term in the energy-momentum tensor gives 
the time-independent contribution 
$3M_P^2H_0^2\,\Omega_{\Lambda, 0}$
to \eqref{rhoDE}. The spatial derivative 
term gives a contribution with the relative 
factor $2\widetilde\epsilon (1+z)^2$. 
In terms of the model parameters, 
$\Omega_{\Lambda, 0}$ is given by\footnote{%
The fractional energy density of dark energy at
present is the sum of the two terms, $\Omega_{\rm DE, 0}=
\left(1+2\widetilde\epsilon\right)\Omega_{\Lambda, 0}$,
but since $\widetilde{\epsilon}\ll 1$, one can take
$\Omega_{\rm DE, 0}\approx\Omega_{\Lambda, 0}$, in practice.} 
\al{
	\Omega_{\Lambda, 0}=\frac{1}{6}\tilde{c}_*\frac{H_A^4}{H_0^2M_P^2}\left(\frac{m_0}{m_A}\right)^2
	\,,
}
with $\tilde{c}_*=c_* (H_I/H_A)^{2\epsilon}$, where 
$c_*$ depends on the size 
of the critical bubble\footnote{%
We have $c_*=3/8\pi^2$ in the small bubble limit,
and $c_*=3/4\pi^2$ when the bubble occupies half
of the ancestor de Sitter space.}~\cite{Aoki:2017scu}. 
We can take $(H_I/H_A)^{2\epsilon}\sim 1$ 
when $\epsilon$ is sufficiently small.
The parameters $H_A$, $m_A$, $m_0$ in the model are related
to the observables, $\Omega_{\Lambda,0}$ and $\xi$ as 
\al{
	&{H_A\over M_P}=\sqrt{6 \xi \Omega_{\Lambda,0} \over 
	\tilde{c}_{*}c_{\epsilon}}
	\,,\ \
	{m_A\over M_P}=\sqrt{\frac{6\Omega_{\Lambda,0}}{\tilde{c}_{*}}}\,
\frac{\xi}{c_{\epsilon}}\,\frac{m_0}{H_0}
	\,.
\label{HAmA}
}
We will forecast the constraints for 
$\xi$, rather than $\tilde{\epsilon}=\xi\Omega_{K,0}$. 
Since the definition of $\xi$ is independent of 
$\Omega_{\rm K,0}$, it is not constrained to be very small;
we expect $\xi ={\cal O}(1)$ as a natural choice.
If we can determine $\xi$ 
from observations, \eqref{HAmA} allows us to determine an
important parameter $H_A$.

\paragraph*{Fisher analysis for galaxy surveys.---}
 SKA~\cite{SKA} is a ground based array of radio telescopes which
covers about 3/4 of the sky and observes galaxies
by detecting the 21cm emission line of neutral hydrogen. 
Euclid~\cite{Euclid} is a
satellite based telescope working in the visible and 
near-infrared wavelength domains. 
SKA phase 2 (SKA2) and Euclid are both expected to 
observe a billion galaxies up to the redshift $z\sim 2$.
In our analysis, we use the survey specifications 
described in \cite{Yahya:2014yva} for SKA, and in
\cite{Amendola:2016saw} for Euclid.
We take the power spectrum of galaxy distribution 
as an observable, and forecast the constraints
on the parameters,  
following the standard 
procedure (see e.g., \cite{Tegmark:1997rp, Seo:2003pu, 
White:2008jy, Yamauchi:2016ypt}).

As the fiducial cosmological model, we take a $w$CDM model 
whose dark energy is characterized by $w$ in \eqref{wz}
with the negative spatial curvature. 
The Hubble parameter $H(z)$ obeys
\al{
&\frac{H^2}{H_0^2}
=\Omega_{\rm m,0}(1+z)^3
+\widetilde\Omega_{\rm K,0}(1+z)^2
+\Omega_{\Lambda,0}
\,,
\label{H2H0}\\
&\frac{\dot H}{H_0^2}=-\frac{3}{2}\Omega_{\rm m,0}(1+z)^3
-\widetilde\Omega_{\rm K,0}(1+z)^2
\,.
\label{dotH}
}
Interestingly, the spatial derivative term in the 
energy-momentum tensor contributes exactly in the same way
as the spatial curvature to Eqs.~\eqref{H2H0} and \eqref{dotH},
effectively replacing $\Omega_{\rm K,0}$ with 
$\widetilde\Omega_{\rm K,0}\equiv\left(1+2\xi\Omega_{\Lambda,0}
\right)\Omega_{\rm K,0}$. Thus there is a tendency for 
degeneracy between $\Omega_{\rm K,0}$ and $\xi$. 
On the other hand, the angular diameter distance, 
$D_{\rm A}(z)=\left((1+z)H_0\sqrt{\Omega_{\rm K,0}}
\right)^{-1}
\sinh\left(H_0 \sqrt{\Omega_{\rm K,0}}\int_0^z
H^{-1}(z^\prime)\dd z^\prime\right)$,
is defined in terms of the true curvature 
$\Omega_{\rm K,0}$. This fact is expected to break
the degeneracy.

We consider a model for galaxy distribution in the 
linear regime. The matter density contrast 
$\delta_{\rm m}$ satisfies the $k$-independent equation
at the linear level (see e.g., \cite{Peebles}),
\al{
	\ddot\delta_{\rm m} +2H\dot\delta_{\rm m} -\frac{3}{2}H^2\Omega_{\rm m}\delta_{\rm m} =0
	\,.
}
An object of interest is the linear growth rate
$f\equiv \frac{\dd\ln\delta_{\rm m}}{\dd\ln a}$.
The observed galaxy power spectrum in the redshift space is well described by
\al{
	P_{\rm g}({\bm k};z)
		=\left( b(z)+f(z)\mu^2\right)^2 P_{\rm m}(k;z)
		e^{-k^2\mu^2\sigma_{\rm NL}^2}
	\,,
}
where $P_{\rm m}(k,z)$ is the linear matter power spectrum, 
and $b(z)$ is the so-called galaxy bias function,
which should be chosen according to the type of target
galaxies for the particular observation: 
for SKA, $b(z)=c_1\exp(c_2 z)$ with constant $c_1$
and $c_2$; for Euclid, $b=\sqrt{1+z}$. 
The term $f\mu^2$, where $\mu$ is the cosine of the
angle between the line of sight and the wave vector ${\bm k}$,
represents the redshift space distortion~\cite{Kaiser:1987qv}, 
due to the contribution 
to the observed redshift from the peculiar velocity driven by the
clustering of matter (making dense region look denser). 
The factor $e^{-k^2\mu^2\sigma_{\rm NL}^2}$
with a free parameter $\sigma_{\rm NL}\approx 7\,[{\rm Mpc}]$ 
to be marginalized, 
is introduced to represent the inaccuracies in the observed redshift, 
which results in the line-of-sight smearing. Further, taking into 
account the geometrical effects due to the difference between
the possibly incorrect reference cosmology and the true 
one, the so-called Alcock-Paczynski (AP) effect~\cite{Alcock:1979mp},
the observed power spectrum is given by~\cite{Seo:2003pu}
\al{
	P_{\rm obs}({\bm k}_\perp^{\rm ref},k_\parallel^{\rm ref};z)
		=\left(\frac{D_{\rm A}^{\rm ref}(z)}{D_{\rm A}(z)}\right)^2\frac{H(z)}{H^{\rm ref}(z)}P_{\rm g}({\bm k}_\perp ,k_\parallel ;z)
\nonumber
	\,.
}
Here the angular diameter distance $D_{\rm A}$ and the Hubble parameter $H$ in the reference cosmology 
are distinguished by the subscript `ref', while those in true cosmology have no subscript. 
The wave number across and along the line-of-sight in the two cosmologies are related through
$k_\parallel =\frac{H}{H^{\rm ref}}k_\parallel^{\rm ref}$ and 
${\bm k}_\perp =\frac{D_{\rm A}^{\rm ref}}{D_{\rm A}}{\bm k}_\perp^{\rm ref}$. 

\paragraph*{The forecast.---}
The Fisher matrix for a set of parameters $\{\theta^\alpha\}$ is defined
as $F_{\alpha\beta}=\ave{{\partial\over\partial\theta_\alpha}{\partial\over\partial\theta_\beta}
\log {\cal L}}$ where ${\cal L}$ is the likelihood function
when we regard $\{\theta^\alpha\}$ as probabilistic variables
which depends on the data set. 
The matrix $F_{\alpha\beta}$
is the inverse of the covariance matrix,
and the minimum 1$\sigma$ error on $\theta_\alpha$
is given by $\sqrt{(F^{-1})_{\alpha\alpha}}$ (no sum on $\alpha$). 
In terms of galaxy power spectrum, the Fisher matrix
can be written as~\cite{Seo:2003pu,Tegmark:1997rp}
\al{
	F_{\alpha\beta}&=\sum_{z_i}\int_{k_{\rm min}}^{k_{\rm max}}\frac{\dd^3{\bm k}}{(2\pi )^3}V_{\rm eff}({\bm k};z_i )
\nonumber\\
&\qquad
\times
\frac{\pd\ln P_{\rm obs}({\bm k};z_i)}{\pd\theta^\alpha}\frac{\pd\ln P_{\rm obs}({\bm k};z_i)}{\pd\theta^\beta}
	\,,
}
where the effective volume of the survey is given by 
\al{
	V_{\rm eff}({\bm k};z_i)
		\approx\left(\frac{n_{\rm g}(z_i)P_{\rm obs}({\bm k};z_i)}{n_{\rm g}(z_i)P_{\rm obs}({\bm k};z_i)+1}\right)^2 V_{\rm survey}(z_i)
	\,.
}
The survey volume is divided into bins with the width $\Delta z=0.1$ in the redshift. 
Here $V_{\rm survey}(z_i)$ is the comoving volume of the redshift slice centered at $z_i$.
The minimum wavelength is 
$k_{\rm min}(z_i)=2\pi /V_{\rm survey}^{1/3}(z_i)$.
The maximal wavelength is taken to be the scale beyond which
non-linearities become non-negligible. It is estimated as
$k_{\rm max}(z_i)=0.14(1+z_i)^{2/(2+n_{\rm s})}\,[{\rm Mpc}^{-1}]$.

\begin{figure}[t]
\includegraphics{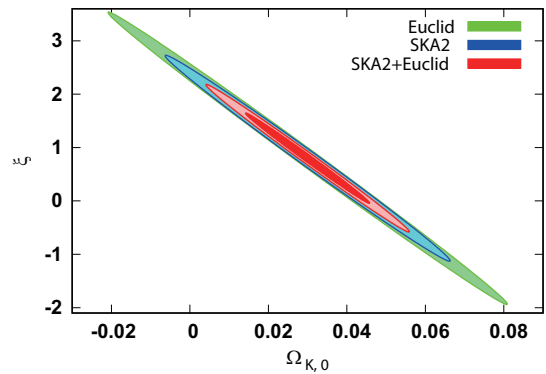}
\caption{Forecast 1$\sigma$ and 2$\sigma$ marginal contours
with the fiducial values
$(\Omega_{\rm K,0}, \xi)=(0.03, 0.8)$
for 
SKA2 galaxy survey (blue), Euclid (green), and the combined analysis (red).}
\label{Fig-forecast}
\end{figure}
Our forecast is performed for the parameter set:
$\{ h,\Omega_{\Lambda,0},\Omega_{\rm K,0}, h^2\Omega_{\rm b,0},\sigma_8,n_{\rm s},\sigma_{\rm NL},b(z_i)\}$
and $\{\xi\}$. 
The fiducial values are taken as
$h=0.67,\ h^2\Omega_{\rm b,0}=0.0222,\ \Omega_{\Lambda,0}=0.6817,\ \sigma_8=0.834,\ n_{\rm s}=0.962,\ \Omega_{\rm K,0}=0.03,$
and $\xi=0.8$.  

FIG.~\ref{Fig-forecast} shows the forecast 1$\sigma$
and 2$\sigma$ contour in the $(\Omega_{\rm K,0}, \xi)$
plane for SKA2 and Euclid.
There is degeneracy between $\Omega_{\rm K,0}$ 
and $\xi$ along the $\widetilde\Omega_{\rm K,0}=\text{const}$.\ 
direction, for the reason mentioned above.
However, for this set of fiducial values of the parameters,
one can see the difference from $\xi=0$ 
(i.e., time independent 
cosmological constant) at the 1$\sigma$ level.

\paragraph*{Discussion.---}
The above choice of parameters, 
$(\Omega_{\rm K,0}, \xi)=(0.03,0.8)$,
is barely consistent with the current constraint
on the spatial curvature. These parameters may
take smaller values. In that case, it might be
difficult to determine these parameters solely 
from the results of galaxy surveys in the near 
future.

To break the degeneracy between $\Omega_{\rm K,0}$
and $\xi$,
it is highly important to combine galaxy surveys with
other observations of spatial curvature which have
different dependence on the angular/luminosity/comoving 
distance. Possible detection of negative curvature by 
observations of the CMB will greatly improve the chance
of determining the parameters, 
since the CMB involves larger $z$ than 
the galaxy surveys, and will be more sensitive to the
angular diameter distance. 
On the other hand, positive curvature at the level of
$\Omega_{\rm K,0}\lesssim -10^{-4}$ 
would falsify bubble nucleation,
as argued in \cite{Kleban:2012ph}, thus its
detection would rule out our model. 

If the results of galaxy surveys are consistent
with our model with non-zero $\xi$, we may wonder
whether this rules out other models. 
For instance, there is a quintessence model, called 
``scaling freezing'' model\footnote{%
It is defined by the potential 
$V(\phi)=V_1 e^{-\lambda_1\phi/M_{P}}
+V_2 e^{-\lambda_2\phi/M_{P}}$ where 
with $\lambda_1>>1$ and $\lambda_2\lesssim 1$. 
Its EoS parameter is well approximated by 
$w_{\rm DE}(a)=-1+1/(1+(a/a_t)^{1/\tau})$ 
with $\tau=0.33$.},
whose EoS parameter approaches $w_{\rm DE}\to -1$ in the future
and $w_{\rm DE}\to 0$ in the past, and the transition occurs
around $a\sim a_t$. Although it would be difficult for
galaxy surveys at $z\lesssim 2$ to distinguish the
past asymptotic values $w_{\rm DE}\to -1/3$ and 
$w_{\rm DE}\to 0$, in fact, 
there is already a strong constraint
for the latter~\cite{Chiba:2012cb, Durrive:2018quo}:
The existing data of CMB, BAO, SNe, 
suggest that the transition has to occur 
quite early 
$a_t<0.11$ (i.e., $z>8.1$). Thus, the model with the past
asymptotic behavior $w_{\rm DE}\to 0$ should have 
$w_{\rm DE}\approx -1$ at $z\lesssim 2$, and cannot be 
responsible for the possible deviation from $w_{\rm DE}=-1$ 
discussed in this Letter.

\paragraph*{Acknowledgements.---}
We thank Jean-Baptiste Durrive for explaining their 
work~\cite{Durrive:2018quo}.
YS would like to thank Lenny Susskind, Steve Shenker,
Bob Wagoner, Adam Brown, Alex Maloney and Yusuke Yamada
for comments at the seminar at Stanford Institute for
Theoretical Physics. 
This work is supported in part by Grants-in-Aid for Scientific
Research (Nos.\ 16K05329, 17K14304) from
the Japan Society for the Promotion of Science.
DSL and CPY are supported in part by the Ministry of
Science and Technology, Taiwan.



\end{document}